\documentclass[manuscript,screen]{acmart}

\usepackage{graphicx}
\usepackage{booktabs}
\usepackage{tabu}

\AtBeginDocument{%
  \providecommand\BibTeX{{%
    \normalfont B\kern-0.5em{\scshape i\kern-0.25em b}\kern-0.8em\TeX}}}

\setcopyright{acmlicensed}
\copyrightyear{2023}
\begin{document}

\title{Better Call GPT, Comparing Large Language Models Against Lawyers}


\author{Lauren Martin, Nick Whitehouse, Stephanie Yiu, Lizzie Catterson, Rivindu Perera}
\affiliation{%
  \institution{AI Center of Excellence, Onit Inc.}
  \city{Auckland}
  \country{New Zealand}}
\email{lauren.martin@onit.com}

\renewcommand{\shortauthors}{Onit AI Center of Excellence}

\begin{abstract}
This paper presents a groundbreaking comparison between Large Language Models (LLMs) and traditional legal contract reviewers—Junior Lawyers and Legal Process Outsourcers (LPOs). We dissect whether LLMs can outperform humans in accuracy, speed, and cost-efficiency during contract review. Our empirical analysis benchmarks LLMs against a ground truth set by Senior Lawyers, uncovering that advanced models match or exceed human accuracy in determining legal issues. In speed, LLMs complete reviews in mere seconds, eclipsing the hours required by their human counterparts. Cost-wise, LLMs operate at a fraction of the price, offering a staggering 99.97 percent reduction in cost over traditional methods. These results are not just statistics—they signal a seismic shift in legal practice. LLMs stand poised to disrupt the legal industry, enhancing accessibility and efficiency of legal services. Our research asserts that the era of LLM dominance in legal contract review is upon us, challenging the status quo and calling for a reimagined future of legal workflows.

\end{abstract}




\maketitle

\section{Introduction}
\label{sec:introduction}

The integration of Artificial Intelligence (AI) into the legal sector has opened a new frontier in legal services. However, as of the current state of research, there appears to be a significant gap in exploratory and experimental studies specifically addressing the capabilities of Generative AI and Large Language Models (LLMs) in the context of determination and discovery of legal issues. Such studies would be instrumental in understanding how these advanced AI technologies manage the intricate task of accurately classifying and pinpointing legal matters, a domain traditionally reliant on the deep, contextual, and specialised knowledge of human legal experts. 

To address the identified gap in the research landscape, this study proposes an experimental and exploratory analysis of the performance of LLMs in the legal domain. The research aims to evaluate the capabilities of LLMs contrasting their performance against human legal practitioners on high volume real-world legal tasks. These types of high volume legal tasks are frequently outsourced or pushed to less experienced lawyers, and given the rapid advancements made by LLMs, raises the question of whether LLMs have achieved a level of legal comprehension that is comparable to the quality, accuracy and efficiency of Junior Lawyers or outsourced legal practitioners on such tasks.

Specifically, we focus on three primary research questions:




\textbf{Do LLMs outperform Junior Lawyers and Legal Process Outsourcers in determination and location of legal issues in contracts?}

This question aims to assess the precision and recall of LLMs compared to human professionals in determining and locating legal issues.

\textbf{Can LLMs review contracts faster than Junior Lawyers and Legal Process Outsourcers?}

In this question, our focus is on evaluating the efficiency (as measured on a temporal perspective) of LLMs in processing and responding to legal queries compared to the time taken by human lawyers.

\textbf{Can LLMs review contracts cheaper than Junior Lawyers and Legal Process Outsourcers?}

This question focuses on evaluating the comparative costs between Lawyers and LPOs to determine whether LLMs are more cost effective.

Through this research, we aim to contribute a comprehensive understanding of the potential capabilities and limitations of LLMs in the legal domain, providing valuable insights for Legal and AI practitioners.

\section{Related Work}

The related work in this area encompasses a broad spectrum of technological advancements in generative AI applied on legal contract reviews and generation of legal contracts. In this section we will be exploring direct and also indirect advancements towards application of Natural Language Understanding (NLU) and Natural Language Generation (NLG) in legal domain giving the priority to generative AI approaches.

Guha et al. \cite{guha2023legalbench} introduce the LegalBench which is a collaborative benchmark for measuring legal reasoning in LLMs. Their work focuses on establishing a benchmark for evaluating the capabilities of LLMs in legal reasoning. By employing the Issue, Rule, Application, Conclusion framework (IRAC), which is a widely used method for organising  legal analysis, the authors have initiated a collaborative project aimed at developing a comprehensive and open-source benchmark. This endeavour emphasises the potential for computational tools to enhance legal practices, especially in administrative and transactional settings. However, the primary objective of LegalBench is not to replace legal professionals with computational systems, but to assess the extent to which current models can support and augment legal reasoning.

Our research, in contrast, delves into a more specialised commercial domain within legal technology, focusing on the application of  LLMs in determining and locating legal issues in contracts. This specific focus on contract analysis sets our work apart from the broader legal reasoning perspective adopted in LegalBench. While Guha et al. emphasise the need for collaborative efforts and in developing benchmarks for legal reasoning with open source data, our study provides a detailed and practical examination of the capabilities of LLMs compared directly with legal practitioners, including Junior Lawyers and LPOs. This aspect of direct comparison is particularly significant as it sheds light on the practical effectiveness and efficiency of LLMs in real-world legal tasks, a perspective not deeply explored in LegalBench.

The significance of our work is further accentuated by its emphasis on the efficiency of LLMs in contract review, a crucial aspect in legal practice where time is a critical factor. Our study not only benchmarks the performance of LLMs against human practitioners but also evaluates their efficiency in terms of time and cost, offering valuable insights into the practical application of these models in legal settings. This focus on both performance and efficiency provides a pragmatic perspective that has an immediate applicability in legal tasks such as contract review, which are common and often time intensive.

In conclusion, while LegalBench establishes a foundational framework for evaluating LLMs in legal reasoning, our research offers a more focused and application-oriented analysis within the legal field. By comparing the capabilities and efficiency of LLMs with those of human legal practitioners in the specific context of contract review, our study addresses a critical gap in the current research landscape. This approach not only benchmarks the effectiveness of LLMs in a practical legal task but also provides substantial contributions to the evolving field of legal technology and its applications.

Tan et al. \cite{tan2023chatgpt} bring a more practical dimension to the theory oriented approach we noticed in the previous research. Although this study does not focus on a deep statistical analysis, it provides some important aspect of practical usage of existing tools. A key aspect of this research, the comparative analysis between ChatGPT and JusticeBot, provides a nuanced understanding of the strengths and limitations of AI-driven legal tools. The study acknowledges that while ChatGPT may not always provide perfectly accurate or reliable information, it offers a powerful and intuitive interface for general public, a significant consideration in enhancing access to legal information.

However, Tan et al. also identify potential areas for improvement, particularly in the reliability and depth of the legal information provided by ChatGPT. In the legal field, where incorrect information can have serious consequences, these aspects are of utmost importance. The research suggests the need for continuous improvement and testing of ChatGPT, especially in the terms of keeping the model updated with the latest legal development and training it on diverse legal scenarios. Although, Open AI claims that ChatGPT currently performs in the level of a legal graduate, there is no evidence in its performance in various legal scenarios. This is one of the key aspect which we tried to address in our research.

Callister \cite{callister2023generative} delves into the philosophical and cognitive implications of AI in law, addressing concerns about the reliability and consistency of AI, and highlighting potential issues such as information hallucination. It also emphasises the transformative nature of AI in shifting the traditional paradigms of legal research. 

While Callister \cite{callister2023generative} focus on the philosophical aspect of AI and LLM's ability to involve in the legal decision making, we tried to offer a viewpoint of practical capabilities and efficiency of LLM's in specific legal tasks. We, however, agree with the cautionary perspective on the broader implications of generative AI for legal epistemology brought to light by Callister \cite{callister2023generative} which needs to be accounted when using generative AI widely in legal decision making. 

Choi et al. \cite{choi2023chatgpt} bring a new dimension to the LLMs applicability in legal domain by focusing on ChatGPT's applicability in legal education. Although this research does not directly align with our study's interest in AI's application in the legal domain, it brings some new perspectives on how LLMs perform in the legal domain. One of the key findings in this study is the analysis of limited ability of ChaptGPT to perform complex legal analysis and issue spotting which we mainly focused on our research. Our research also extends this dialogue by comparing LLMs' performance not just to academic standards (as noticed in \cite{guha2023legalbench}), but also against real world legal practitioners, particularly Junior Lawyers and LPO professionals. This comparison offers a novel perspective on the utility of LLMs in practical legal work, a dimension not fully explored in any of the aforementioned research attempts.

\section{Methodology}

This section focuses on the research methodology covering how data was collected, analysed, prepared, and how the models were selected to be included as part of this research.

\subsection{Overall Design}
The experiment was designed to compare Junior Lawyers and LPOs against LLMs using Senior Lawyers as a benchmark, establishing ground truth for determining and locating legal issues in contracts. This replicates  the process lawyers go through in contract reviews. The aforementioned benchmark formed the basis to assess LLMs, LPOs and Junior Lawyers in their accuracy of making determinations and locating the same legal issues.

\subsection{Ethics Policy}

This research adhered to the ethical standards defined by Onit Inc which covered data collection, analysis, and participant involvement in carrying out primary and secondary research.

All participants were fully informed about the research purpose, data usage, and their rights, including the right to withdraw at any time. In addition, we also ensured the anonymity of participants by replacing identifiable information in the collected data.

All contract data used in the research was anonymised during the data analysis phase to eliminate identifiable details of the involved parties from the process. However, the participant data were de-identified to help further analysis while maintaining the data privacy standards.

We also established ethical oversight and compliance through an involvement of an ethics committee to ensure adherence to the organisation's ethic policies. This included audits of the research process and legal compliance to assure the research activities complied with the applicable data protection and privacy laws.

\subsection{Data Collection and Analysis}

This section will discuss the methods for preparing data for the use in this research and how ground truth was established for the purpose of benchmarking. 

\subsubsection{Dataset Preparation}

Data for this study were collected from ten procurement contracts, sourced from real-world legal agreements and anonymised to preserve confidentiality.

The research used procurement contracts, as they are frequently among the most voluminous types of contracts reviewed by legal practitioners, rivaling the prevalence of Non-Disclosure Agreements (NDAs). The study deliberately excluded NDAs from this analysis due to their typically concise format, which does not provide a complex enough framework to carry out an in-depth evaluation of a LLMs capabilities in determining and locating legal issues.

The jurisdictional scope of the selected contracts was deliberately balanced between the United States (US) which is uniquely based on a combination of statutory and common law, and New Zealand (NZ) which is based on common law. This approach ensures that the findings of our research have relevance across different legal systems, which increases the utility and applicability of the research.

In the dataset of contracts utilised for this study, we took careful measures to preserve confidentiality and data integrity. All references to party names were substituted with pseudonyms, and any sections of the text that required redaction were supplanted with plausible, yet fictitious, information. This methodological step was essential to maintain the authenticity of the contract structure while safeguarding sensitive information, thereby upholding ethical research standards without compromising the quality of our analysis.

Our analysis was structured around a two-factor approach. The first factor comprised a document scenario that provided context, including jurisdiction, historical background, organisational size, and the product or service involved for each party in the contract. The second factor was a set of standardised checks using a contract review playbook, a common practice among legal professionals. The review playbook employed in our study was derived from real-world examples to ensure practical relevance.

Each contract was reviewed from the perspective of a third-party observer, with an equal number of agreements analysed from the vantage points of the supplier and the buyer, respectively. This balanced perspective is designed to simulate the thoroughness and objectivity a legal practitioner would bring to contract analysis, thereby testing the LLMs' performance in a realistic setting.

\subsubsection{Ground Truth Establishment}

Senior Lawyers were instructed to evaluate each contract for the purpose of establishing ground truth data. Their task was to ascertain whether the contract adhered to or deviated from the predefined standards. Additionally, they were required to locate the specific sections of the contract that influence their judgments. In instances where a standard was not met due to the absence of relevant information in the contract, Senior Lawyers were directed to record this explicitly at the conclusion of their assessment for each contract.

This collected data was then aggregated to formulate benchmarks for each evaluative criterion, relying on the majority consensus to determine both compliance with the standard and the precise location of the relevant contractual reference.

Furthermore, Senior Lawyers were requested to record the duration of each contract review. This measure aimed to capture the average time expended by legal professionals in reviewing contracts. This temporal data was intended for subsequent comparison with the time taken by Junior Lawyers, LPOs and LLMs.

\subsection{Establishing hourly rates and LLM costs}
The hourly rates for lawyers were based on inside counsel rates determined through industry benchmark reports such as ACC's 2023 Law Department Compensation Survey \cite{acc_compensation_survey} and external counsel rates determined by market data held by Onit Inc. The costs for LLMs were determined through commercial pricing provided by the service suppliers. 

\subsection{Model Analysis}
The basis for model selection in this research was decided on several factors; prominent models that have emerged in the LLM space, preliminary tests demonstrated promising model capability in our research questions and the context window limits of the models.

\subsubsection{Preliminary Research}
\label{sec:preliminary-research}

The advent of foundational LLMs in mainstream applications has led to the entry of several major competitors in this domain, including OpenAI \cite{openai2023gpt}, Google \cite{anil2023palm}, Anthropic \cite{anthropic_claude}, Amazon \cite{amazon_titan}, and Meta \cite{touvron2023llama}. In selecting the models for our analysis, we focused on those developed by these organisations, as they represent the most prominent entities in this competitive landscape.
    
Preliminary evaluations were carried out on models from these identified companies to assess their applicability and effectiveness within the legal domain. These evaluations involved processing sample contracts through the models, and conducting careful analysis of the outputs. The primary focus of these tests was to analyse the models' reasoning capabilities and their ability to determine legal issues that existed in contracts, as well as their proficiency in pinpointing the location of these issues within these documents.

In developing our research methodology, we placed significant emphasis on determining the optimal size of the context window to effectively address our research questions. The context window's size is a pivotal factor as it directly impacts the ability of the language models to process and analyse the contract documents comprehensively. To ensure a thorough review, we considered the average length of procurement contracts, the complexity of document scenarios required to provide a meaningful context, and the detail necessary in the document checklists.

Our primary goal was to select models capable of handling an extensive amount of contextual information to facilitate accurate and nuanced analysis. As a result, we established a minimum threshold for the context window, and models that could not support at least 16,000 tokens were considered inadequate for our purposes and were subsequently excluded from our analysis.

During our preliminary testing, it was found that models with context windows smaller than 16,000 tokens, such as LLaMA2 and Amazon Titan, required us to split the documents into multiple parts. This segmentation led to a significant portion of the context window being dedicated to repeatedly referencing vital elements, including the document scenario, the review playbook, and the contract's definitions section. This repetition was necessary to ensure that each segment was properly contextualised, but it was also inefficient and detracted from the models' ability to analyse the contract as a whole.

Although there are various techniques, such as Retrieval Augmented Generation (RAG), that mitigate the limitations of smaller context windows, our preliminary evaluation indicated that these methods did not measure up to the performance of models with larger context windows. The RAG approach, in particular, was found to introduce discontinuities in the contextual understanding of the contracts. These gaps were significant enough to compromise the models' performance, making them unsuitable for an accurate comparison with the level of comprehension and analysis expected of a human lawyer. Consequently, our study focused on models with large enough context windows to ensure representative assessment of their capabilities in legal contract analysis.

\subsubsection{Model Settings}
Table \ref{tab:model_settings}

\begin{table}[ht]
\begin{center}
    \begin{tabu} {X[l] X[r] X[r] X[r]}
         \toprule
         \textbf{Model} &  \textbf{Temperature} &  \textbf{Token Input} & \textbf{Training Data}\\ 
         \midrule
         GPT4 32k &  0.2 &  32,768 & Up to Sept 2021\\ 
         GPT4-1106 &  0.2 &  128,000& Up to Apr 2023\\  
         GPT3.5-turbo-16k & 0.2& 16,385& Up to Sept 2021\\ 
        Claude 2.1 & 0.2& 200,000& Up to early 2023\\ 
        Claude 2.0 & 0.2& 100,000& Up to early 2023\\  
        Palm2 Text-bison-32k-2 & 0.2 & 32,768 & Up to mid 2021\\ 
        \bottomrule
\end{tabu}

\end{center}
\caption{Model Settings}
    \label{tab:model_settings}
\end{table}




\subsection{Prompt Engineering}
Each LLM had a different set of requirements for the prompts needed to complete the contract review tasks. Prompts were structured with the following elements; the role, the task and the context. The role required the LLMs to adopt the persona of a lawyer in the task it was to undertake. The LLMs were tasked with  reviewing contracts against given standards by determining and locating where those issues were in contracts. The context element was designed to mirror the instructions typically provided to a Lawyer, LPO or Contract Reviewer. Context included a range of factors such as the target audience for the contract, pertinent background information regarding the contracting parties, and the specific scenario under which the contract was being negotiated. Examples of the prompts used for each LLM can be referred to in Tables \ref{tab:system_prompt_GPT_4} and \ref{tab:system_prompt_claude 2.1}.

\section{Evaluation and Results}

\subsection{Legal practitioner comparison}

\subsubsection{Inter-annotator Agreement}

To assess the level of agreement among the three groups of legal practitioners (Senior Lawyers, Junior Lawyers, and LPOs), we conducted an inter-annotator agreement analysis employing Cronbach's Alpha \cite{cronbach1951coefficient} as the primary method of evaluation. 

\begin{figure}[h]
\centering
\includegraphics[width=0.5\textwidth]{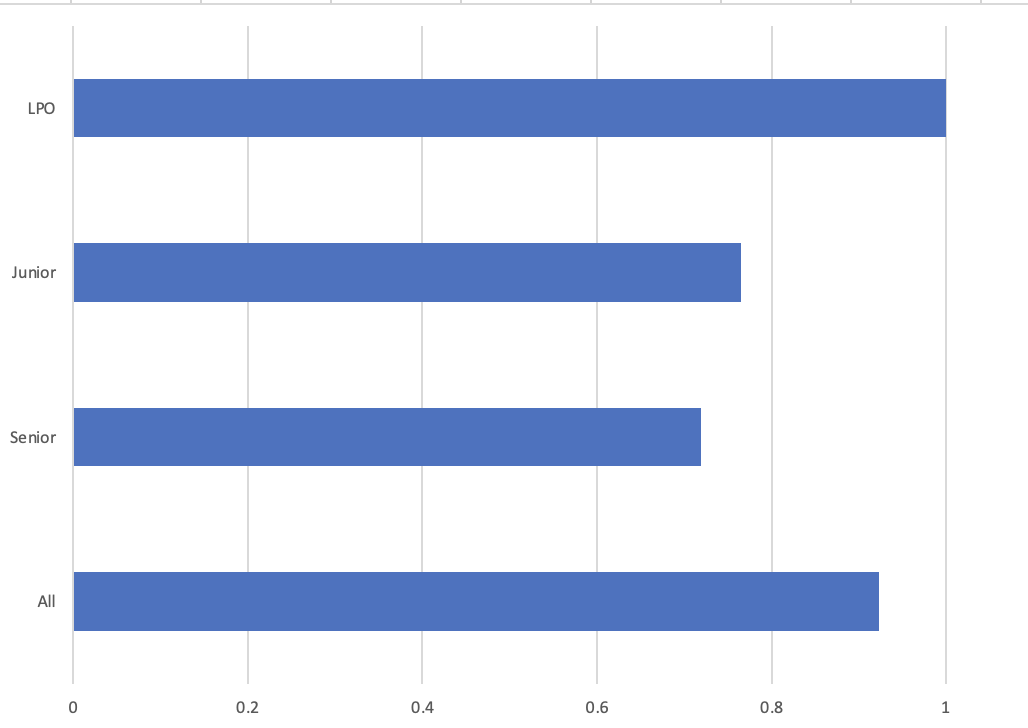}
\caption{Level of agreement on issues by role}
\label{fig:inter-annotator}
\end{figure}

The inter-annotator agreement results as shown in Fig. \ref{fig:inter-annotator} revealed noteworthy findings in our research on the applicability of LLMs in the legal domain. The overall agreement among all participants was remarkably high, with an Alpha value of 0.923366, indicating a very strong consensus. Interestingly, the agreement among Senior Lawyers alone was the lowest at 0.719308, suggesting a more varied approach to issue identification in contracts among experience practitioners. Junior lawyers demonstrated a slightly higher level of agreement, with an Alpha value of 0.765058, potentially reflecting more consistent training methodologies or adherence to established legal frameworks. Notably, the agreement among LPO practitioners reached a perfect Alpha of 1, indicating absolute uniformity in their responses. 

\subsection{Comparing LLMs against Junior Lawyers and LPOs}

\subsubsection{Performance Comparison}

To address the research questions outlined in Section \ref{sec:introduction}, our research centered on conducting a comparative analysis of the accuracy of different LLMs (as detailed in Section \ref{sec:preliminary-research}) in relation to Junior Lawyer and LPOs. This comparison was conducted with the Senior Lawyers' determinations serving as the ground truth data against which those results were assessed.

\begin{table}[ht]
    \begin{center}
        \begin{tabu}{X[2] X[1,l] X[1,r] X[1,r] X[1,r]}
            \toprule
             &    \textbf{Precision} & \textbf{Recall} & \textbf{F-score} & Loss\\
             \midrule
           LPO &    0.933 & 0.823 & 0.874 & 0.126 \\ 
           GPT4-1106 &    0.835 & 0.910 & 0.871 & 0.129\\
           Junior &    0.876 & 0.845 & 0.860 & 0.140\\
           Claude 2.0 &    0.743 &0.907 & 0.817 & 0.183\\
           GPT4 -32k &    0.958 & 0.723 & 0.820 & 0.180\\ 
         Claude 2.1 &    0.723 & 0.917 & 0.809 & 0.191\\ 
         Palm2 text-bison &    0.617 & 0.831 &0.708& 0.292\\ 
         GPT3.5 &    0.531&0.864 & 0.657 & 0.343\\
         \bottomrule
            
        \end{tabu} 
    \end{center}
    \caption{Performance ranked by F-score in determining legal issues}
    \label{tab:my_label}
\end{table}



\begin{table}[ht]
    \begin{center}
        \begin{tabu}{X[2] X[1,l] X[1,r] X[1,r] X[1,r]}
            \toprule
             &    \textbf{Precision} & \textbf{Recall} & \textbf{F-score} & Loss\\
             \midrule

                 LPO&    0.915 &0.666 &0.770& 0.230\\  
                 GPT4-32k&    0.847 &0.660 &0.740& 0.260\\  
                 Claude 2.1&    0.911 &0.569 &0.701& 0.299\\  
                 GPT4-1106&    0.857 &0.575 &0.686& 0.312\\  
                 Junior&    0.866 &0.542 &0.667& 0.333\\  
                 Claude 2.0&    0.812 &0.517 &0.631& 0.369\\  
                 GPT 3.5&    0.633 &0.607 &0.620& 0.380\\  
                 Palm2 text-bison&    0.563&0.378&0.452& 0.548\\ 
                
         \bottomrule
        
        \end{tabu} 
    \end{center}
    \caption{Performance ranked by F-score in locating legal issues}
    \label{tab:location_fscores}
\end{table}



This comparative evaluation was carried out on two fronts similar to previous evaluations: the ability to determine and locate legal issues in contracts.

In the issue determination aspect, both GPT4-1106 and LPO practitioners emerged as the top performers with an F-score of 0.87, indicating a high level of precision and reliability in this aspect. Junior lawyers achieved an F-score of 0.86 suggesting that their expertise is slightly below the leading LLMs in determining legal issues. Other models, such as Claude 2.0 and GPT4-32k, also showed strong performance with F-scores of 0.82. Whereas GPT3.5 and Palm2 text-bison trailed with lower accuracy scores.

The proficiency in pinpointing legal issues within contracts presented a different competitive scenario. Here, LPOs took the lead with an F-score of 0.77, edging out the top-performing LLM in this metric, GPT4-32k, which achieved a F-score of 0.74. Notably, GPT4-1106, while excelling in issue determination, fell short in localization accuracy, achieving an F-score of 0.69. This divergence underscores the distinct competencies and limitations that are specific to each LLM. Models such as Claude and GPT3.5 exhibited moderate localization capabilities. In stark contrast, Palm2 Text-bison's performance was markedly lower, particularly in issue location, with an F-score of  0.46.

\subsubsection{Time comparison}
Alongside the performance comparison, a time analysis was conducted to evaluate how Junior Lawyers and LPOs compared to LLMs in terms of time efficiency when reviewing legal contracts.

The time efficiency  analysis of each group yielded significant insights in our study as shown in Table \ref{tab:time_comparison}. Amongst the human practitioners, Senior Lawyers demonstrated the most efficiency averaging 43.46 minutes per document, while Junior Lawyers took slightly longer, averaging 56.17 minutes. Remarkably, LPOs took substantially longer with an average of 201 minutes. 

Conversely, all LLMs demonstrated markedly higher time efficiency. The longest processing time was taken by GPT-1106 at an average of 4.7 minutes and the shortest Palm2 text-bison at 0.73 minutes. These findings indicate that LLMs significantly outperform both Junior Lawyers and LPOs in terms of time efficiency for the task of reviewing legal contracts. 

\begin{table}[ht]
    \begin{center}
        \begin{tabu}{X[l] X[r]}
        \toprule
             \textbf{Role} & \textbf{Average Time per Document (Minutes)} \\
             \midrule
             Senior Lawyer & 43.46 \\
             Junior Lawyer & 56.17 \\
             LPO &  201.00 \\
             GPT 4-1106 & 4.70 \\
             GPT 4-32k & 2.11 \\
             GPT 3.5 & 1.44 \\
             Claude 2.1 & 2.05 \\
             Claude 2.0  & 1.63 \\
             Palm2 text-bison & 0.73 \\
             \bottomrule
        \end{tabu}
        
    \end{center}
    
    \caption{Time Comparison}
    \label{tab:time_comparison}
\end{table}

\subsubsection{Cost comparison}

As a part of the evaluation, we also conducted a detailed cost comparison between lawyers, LPO practitioners and LLMs utilised in this research (GPT4-1106, GPT4 32k, GPT 3.5, Claude 2.0, Claude 2.1 and Palm2 Text-bison). This comparison is crucial to understand the economic implications of employing LLMs in the legal domain, especially in the tasks involving the determination and location of legal issues  in contracts. The cost per contract for human practitioners was calculated based on their average time spent on a contact and their respective hourly rates. For LLMs, the cost was determined by the average number of input and output tokens, and the respective cost per token, except for Palm2 text-bison which was priced based on the number of characters used.

The cost comparison, as shown in Table \ref{tab:cost_comparison}, illustrates a stark constrast in cost efficiency between human practitioners and LLMs with LLMs offering a significantly lower cost per document. 

\begin{table}[ht]
    \begin{center}
        \begin{tabu}{X[l] X[r]}
        \toprule
             \textbf{Role} & \textbf{Average Cost per Document (\$USD)} \\
             \midrule
             Senior Lawyer & 75.92 \\
             Junior Lawyer & 74.26 \\
             LPO &  36.85 \\
             GPT 4-1106 & 0.25 \\
             GPT 4-32k & 1.24 \\
             GPT 3.5 & 0.05 \\
             Claude 2.1 & 0.02 \\
             Claude 2.0  & 0.02 \\
             Palm2 text-bison & 0.03 \\
             \bottomrule
        \end{tabu}
        
    \end{center}
    
    \caption{Cost Comparison}
    \label{tab:cost_comparison}
\end{table}

\section{Discussion and Further work}

The findings of this study provide a comprehensive analysis of the performance of LLMs in comparison to Junior Lawyers and LPOs in the context of legal contract review. Our discussion focuses on the three core questions that drive this research.

\subsection{Do LLMs outperform Junior Lawyers and LPOs in determination and location of legal issues in contracts?}
The results show LLMs demonstrate comparable, if not superior, performance in determining legal issues when compared to Junior Lawyers and LPOs. However, a LLMs ability to locate issues within contracts, particularly where a standard is not present, is model-dependent and may not consistently outperform human practitioners, highlighting the importance of selecting the right model for the legal task.

Evidence from our results supports this view, with LLMs, specifically GPT4-1106, achieving an F-score of 0.87 in issue determination, on par with LPO results and slightly better than Junior Lawyers. However, in issue location, GPT4-1106 only achieved an F-score of 0.69 which was lower than the LPOs' score of 0.77, by contrast, the top performing LLM in issue location, GPT4-32k, achieved an F-score of 0.74. 

Arguably, the lower performance by LLMs in locating issues could impact their ability to accurately mark up contracts on their own, constraining their effectiveness in eliminating the role of a Lawyer in contract review. However, with the pace of LLM improvement and the many LegalTech software tools in market utilising a variety of LLMs in combination with more traditional software approaches, we believe the 0.03 F-score difference between LPOs and LLMs is trivial, and we assert that when it comes to accuracy in contract review, LLMs are now operating at the same level of quality as LPOs.

\subsection{Can LLMs review contracts faster than Junior Lawyers and LPOs?}
In evaluating the efficiency of contract review, our analysis, as detailed in Table \ref{tab:time_comparison} indicates a substantial difference in the speed of contract review between LLMs and human reviewers. Specifically, the fastest LLM, Palm2 text-bison completed the contract review tasks in an average of 0.728 minutes. This contrasts sharply with the average time of 56.17 minutes for a Junior Lawyer and 201 minutes for an LPO. Such a disparity not only highlights the superior processing capabilities of LLMs but also suggests a paradigm shift in how contracts can be reviewed.

For instance, the fastest LLM outpaces LPOs by an astonishing 276-fold and Junior Lawyers by 77-fold. Such metrics not only highlight the transformative capabilities of LLMs in legal practice but also illustrate their potential to revolutionize contract review processes. In the time it takes a human reviewer to scrutinize a single contract, an LLM could analyze hundreds. Moreover, LLMs possess the capacity to handle multiple requests concurrently, exponentially increasing the volume of contracts they can assess. This level of efficiency will significantly alter the landscape of legal services, offering unprecedented scalability and speed.

A frequent skepticism surrounding the adoption of LLMs pertains to the initial time investment required for their setup. Our research found that setting up, testing, fine-tuning, and validating the LLMs' system prompts took an average of 16 hours. Notably, this investment is on par with the time dedicated to instructing Junior Lawyers and LPOs for similar tasks. This equivalence in preparatory time challenges the notion that LLMs' speed advantage is offset by their setup requirements, reinforcing the position that LLMs offer significant time-saving benefits once operational.

\subsection{Can LLMs review contracts cheaper than Junior Lawyers and LPOs?}
Our cost analysis, outlined in Table \ref{tab:cost_comparison}, provides a stark economic comparison between LLMs and traditional human contract review. The data reveal that the cost per contract review for LLMs is significantly lower than that for human reviewers. For instance, while a Junior Lawyer incurs an average cost of 74 dollars per contract review, the fastest LLM performed the same task for approximately 2 cents. This equates to a cost reduction of more than 99.97 percent when compared to Junior Lawyers, and 99.94 percent compared to LPOs.

These figures not only reflect the raw efficiency of LLMs but also allude to the potential for scaling legal services. The adoption of LLMs will enable legal departments and law firms to substantially increase their contract review throughput without a proportional increase in costs. This scalability suggests a future where legal services could become more accessible to a wider array of users or clients, revolutionizing the industry. Nevertheless, this optimistic outlook should be tempered with careful consideration of the quality of reviews and the need for ongoing supervision of LLM outputs.

\subsection{Implications for the Legal Industry}
The implications of this research on the legal industry are profound and multifaceted. We believe LLMs are already at a point where they can significantly disrupt LPOs. Our research shows LLMs are as accurate, yet faster and more cost-effective at the tasks LPOs are typically engaged in. There is little doubt that broad adoption of LLMs will lead to a significant shift in the LPO business model, necessitating a need for LPOs to pivot from providing manual legal process services, perhaps to becoming managers and operators of LLM technology, focusing on quality control, interpretation, and nuanced application of the insights generated by these models. 

We anticipate that Junior Lawyers will also encounter disruption, though the nature of these disruptions will differ. While our research does not explore a broad enough spectrum of the capabilities performed by Junior Lawyers to establish a one to one comparison, it does indicate that LLMs are already outperforming Junior Lawyers in specific areas. We believe the partial adoption of LLMs will begin to reduce the demand for entry-level legal positions, as they are currently defined, over time. In contrast, established Junior Lawyers will engage in more complex, high-value work earlier in their careers, as routine tasks are gradually transitioned to LLMs.

The ramifications for legal departments and law firms are unmistakable: LLMs present remarkable efficiency gains and cost savings. Early adopters of this technology are poised to gain a significant competitive edge, likely triggering an arms race in the legal sector. Despite the potential for controversy and litigation, given the regulatory landscape of the legal profession, genuine resistance to LLM adoption may be less about the validity of the technology—as evidenced by our research—and more about protecting established interests. Such opposition could be perceived as a strategic manoeuvre to maintain the status quo rather than a genuine concern for the practical or ethical shortcomings of AI in legal practice, this, of course, would run counter to the industry wide pursuit for greater access to justice, and potentially effect trust in the legal sector.

\subsection{Further Work}
To build on our findings, further research is needed to evaluate LLM performance across a broader range of contract types and to expand the ground truth dataset. Additionally, we are interested in exploring LLMs' capabilities in contract negotiation—a complex area requiring contextual understanding beyond the contract text itself. This future work will help to map the full potential of LLMs in the legal industry and address any limitations observed in the current study.

\section{Conclusion}

Our findings show that LLMs perform on par with LPOs and Junior Lawyers, accurately determining legal issues within contracts. Notably, when it comes to the speed of contract review, LLMs demonstrate a significant advantage due to their computational efficiency, which enables them to process and analyse text much faster than human practitioners. This advantage is a game-changer in terms of productivity and turnaround times for contract review.

Moreover, the cost analysis show that LLMs offer a substantially cheaper alternative for contract review when compared to the costs associated with Junior Lawyers and LPOs. The combination of high accuracy, rapid processing, and lower costs makes LLMs an attractive option for legal practitioners and firms looking to optimise their contract review processes.

In conclusion, the evidence from this study supports the assertion that LLMs are not only viable but are superior tools for legal contract review over Junior Lawyers and LPOs. They can deliver accurate results at a fraction of the time and cost required by traditional human-based review. As the technology continues to advance, it is likely that LLMs will play an increasingly crucial role in the legal industry, reshaping the landscape of legal services and offering new opportunities for efficiency and scalability.

\section{Acknowledgments}

This research was conducted by Onit’s Artificial Intelligence Center of Excellence, with special mention to the principal investigators:

Lauren Martin, Legal Services Manager (PI), Stephanie Yiu, Legal Engineer (Co-PI), Lizzie Catterson, Legal Engineer and Data Analyst (Co-PI), Dr. Rivindu Perera, VP of AI and Data Science (Co-PI), Nick Whitehouse, Co-founder and Managing Director (Co-PI).

\bibliographystyle{abbrv}
\bibliography{sample-base}

\appendix

\section{Appendix}

\subsection{System prompts used in research}

For each LLM used in the research a system prompt was constructed to send with all the relevant information for each contract. Referenced is the system prompt used for GPT 4-1106, see Table \ref{tab:system_prompt_GPT_4}, and Claude 2.1, see Table \ref{tab:system_prompt_claude 2.1}.

\begin{table}[ht]
    \begin{center}
    \begin{tabu} to \textwidth {X[1] X[2]}
        \toprule
        \textbf{Model} & \textbf{System Prompt}\\
        \midrule
        GPT 4-1106 & As an experienced lawyer, your task is to meticulously review contracts currently under negotiation from the perspective of the Supplier party.       

Your intended audience is internal counsel, and the contracts you are reviewing are standalone agreements between the parties. These contracts are a mix of US and NZ documents. Your goal is to identify any legal issues that could potentially affect your client's legal rights or obligations such as language in the contract which does not meet the standard or requirement of the party you represent. These issues could be points of contention, ambiguities, potential risks, or compliance matters that could lead to a dispute or liability.       

Your analysis should be based on the following:       

- The document scenario given for each contract, which provides background context on both parties involved in the agreement including the size, industry, and product/service of the parties and any specific areas of concern in the contract.      

- The provided contract.     

- The provided checklist for the client you represent, which lists the items they deem important in the contract. These items, or 'checks', indicate the legal standard or requirement of the party you represent.        

For each section of the provided contract, carefully analyze it and identify ALL of the checks listed in the provided checklist. Format your answer as a JSON object that includes an array of JSON objects, each representing a different check from the checklist. Each of these JSON objects MUST include the following keys:        

1. \'checklist\_Numbering': The numbering of the check from the checklist. Please STRICTLY adhere to the EXACT numbering format used in the checklist.    

2. \'contract\_Lines': An array of sentences or clauses in the contract that corresponds to the check.  

If the contract does not contain a specific sentence or clause related to the check, then leave this key blank. 

3. 'explanation': A comprehensive and descriptive explanation of how the contract meets/fails to meet the legal standard in the check.     

4. ‘check\_Met’: Indicate whether the check is ‘Met’/‘Not met’. If the contract\_Lines key is 'null', then indicate 'Not met'.   

5. ‘assumptions’: List any additional assumptions that were necessary to answer this check.       

Important:      

- Provide a separate JSON object for EACH check.  

- Ensure each JSON object is formatted in the following order: 'checklist\_Numbering', 'contract\_Lines', 'explanation', 'check\_Met', 'assumptions'. 

- Ensure to adhere strictly to the structure and naming of the keys as described above.  

- Remember to thoroughly reread the contract and treat each check independently for accurate identification.  

- The context provided in the document scenario should guide your review process. \\

        \bottomrule
        
    \end{tabu}
        
    \end{center}
    \caption{System Prompt: GPT 4-1106}
    \label{tab:system_prompt_GPT_4}
\end{table}

\begin{table}[ht]
    \begin{center}
    \begin{tabu} to \textwidth {X[1] X[2]}
        \toprule
        \textbf{Model} & \textbf{System Prompt}\\
        \midrule
Claude 2.1 & Here is a document scenario for you to use as background context for reviewing a contract:

<document\_scenario> XXX </document\_scenario>

Here is a contract for you to review: 

<contract> XXX </contract>

Here is a checklist for you to reference for your task: 

<checklist> XXX </checklist>

[Task instructions] 

You are an experienced lawyer, your task is to meticulously review contracts currently under negotiation from the perspective of the Supplier party. 

Your intended audience is internal counsel, and the contracts you are reviewing are standalone agreements between the parties. These contracts are a mix of US and NZ documents. Your goal is to identify any legal issues that could potentially affect your client's legal rights or obligations such as language in the contract which does not meet the standard or requirement of the party you represent. These issues could be points of contention, ambiguities, potential risks, or compliance matters that could lead to a dispute or liability. 

Your analysis should be based on the following: 

- The document scenario given for each contract, which provides background context on both parties involved in the agreement including the size, industry, and product/service of the parties and any specific areas of concern in the contract. 

- The provided contract.

- The provided checklist for the client you represent, which lists the items they deem important in the contract. These items, or 'checks', indicate the legal standard or requirement of the party you represent.

For each section of the provided contract, carefully analyze it and identify ALL of the checks listed in the provided checklist. 

You ALWAYS follow these guidelines when writing your response: 

<guidelines> 

Format your answer as a JSON object that includes an array of JSON objects, each representing a different check from the checklist. Each of these JSON objects MUST include the following keys: 

1. 'checklist\_Numbering': The numbering of the check from the checklist. Please STRICTLY adhere to the EXACT numbering format used in the checklist. 

2. 'contract\_Lines': An array of sentences or clauses in the contract that corresponds to the check. If the contract does not contain a specific sentence or clause related to the check, then leave this key blank. 

3. 'explanation': A comprehensive and descriptive explanation of how the contract meets/fails to meet the legal standard in the check. 

4. ‘check\_Met’: Indicate whether the check is ‘Met’/‘Not met’. If the contract\_Lines key is 'null', then indicate 'Not met'. 

5. ‘assumptions’: List any additional assumptions that were necessary to answer this check.

Important: 

- Provide a separate JSON object for EACH check. 

- Ensure each JSON object is formatted in the following order: 'checklist\_Numbering', 'contract\_Lines', 'explanation', 'check\_Met', 'assumptions'. 

- Ensure to adhere strictly to the structure and naming of the keys as described above. 

- Remember to thoroughly reread the contract and treat each check independently for accurate identification. 

- The context provided in the document scenario should guide your review process. 

- Provide the JSON response immediately without preamble. 

</guidelines>\\ 

        \bottomrule
        
    \end{tabu} 
    \end{center}
    \caption{System Prompt: Claude 2.1}
    \label{tab:system_prompt_claude 2.1}
\end{table}

\subsection{Sample of Incorrect LLM determinations}

In the course of our research, we observed that emergent LLMs, particularly GPT 4-1106, exhibited a level of accuracy in identifying legal issues that was on par with that of Junior Lawyers and LPOs. A notable finding, as detailed in Table \ref{tab:missed_determinations}, was the trend across these groups showing a preference for precision over recall. This tendency often led to an under-identification of legal issues, marked by a frequent conclusion of non-existence of issues where they actually were present.

This phenomenon was especially pronounced in the LLMs, where a significant portion of inaccuracies was linked to the misinterpretation of specific wordings. The examples in the aforementioned table illustrate these inaccuracies, which were largely due to challenges in grasping the nuances of legal language. In comparison, the Senior Lawyers, serving as the benchmark for this study, showed exceptional care in interpreting subtle differences in language, indicating a sophisticated understanding of semantic variations. This juxtaposition between the performance of LLMs and human experts offers valuable insights into the current capabilities and limitations of LLMs in legal issue determination, highlighting areas for potential improvement and further research.

\begin{table}[ht]
    \begin{center}
        \begin{tabu} to \textwidth {X[1] X[3] X[1] X[2]}

            \toprule
            \textbf{Required Check} &  \textbf{Contract Reference} & \textbf{Ground Truth Response} & \textbf{LLMs Response}\\
            \midrule
            
         Immediate written notice is required in the event of data or security breach &  6.2 Notifications. Each party shall promptly notify the other party of any actual or suspected misuse or unauthorised disclosure of the other party’s Confidential Information.&  Standard is not met as it does not specify the requirement for notice to be written.&The contract requires each party to promptly notify the other party of any actual or suspected misuse or unauthorized disclosure of the other party's Confidential Information. -  GPT 3.5

The contract requires immediate written notice in the event of a data or security breach. - text-bison\\
            & & & \\
         Publicity: Ensure written consent is required for any public announcements or use of logo or branding. &  15.3. Public Announcements. Neither Party shall issue or release any announcement, statement, press release, or other publicity or marketing materials relating to this Agreement or, unless expressly permitted under this Agreement, otherwise use the other Party’s trademarks, service marks, trade names, logos, domain names, or other indicia of source, association, or sponsorship, in each case, without the prior written consent of the other Party, which consent shall not be unreasonably withheld; provided, however, that we may, without your written consent, include or display your name, logo and other indicia in our lists of current or former customers in promotional and marketing materials.&  Standard is not met as not all publicity requires written consent&Section 15.3 states written consent is required for any public announcements or use of logos/branding, meeting this requirement. - Claude 2.1\\

         \bottomrule

        \end{tabu}
        
    \end{center}
    \caption{Examples of Determination missed by LLMs}
    \label{tab:missed_determinations}
\end{table}

\end{document}